\documentclass[twocolumn] {revtex4-1}
\usepackage{amsmath}  % needed for \tfrac, \bmatrix, etc.
\usepackage{amsfonts} % needed for bold Greek, Fraktur, and blackboard bold
\usepackage{graphicx} % needed for figures
\usepackage{braket}
\usepackage{hyperref}
\hypersetup{
	colorlinks=true,
	linkcolor=blue,
	filecolor=magenta,      
	urlcolor=cyan,
}

\begin{document}

% Be sure to use the \title, \author, \affiliation, and \abstract macros
% to format your title page.  Don't use lower-level macros to  manually
% adjust the fonts and centering.

\title{Are we living in Non-Commutative Space?\\
\small{revisiting the classic hydrogen atom system}}

% In a long title you can use \\ to force a line break at a certain location.

\author{Pulkit S. Ghoderao}
\email{pulkit.ghoderao18@imperial.ac.uk}
\affiliation{Theoretical Physics Group, Imperial College London, London SW7 2AZ} % optional
%\altaffiliation[permanent address: ]{101 Main Street, 
 % Anytown, USA} % optional second address
% If there were a second author at the same address, we would put another 
% \author{} statement here.  Don't combine multiple authors in a single
% \author statement.
\author{P. Ramadevi}
\email{ramadevi@phy.iitb.ac.in}
\affiliation{Department of Physics, Indian Institute of Technology Bombay, Powai, Mumbai - 400 076}
% Please provide a full mailing address here.

%\author{David P. Jackson}
%\email{ajp@dickinson.edu}
%\affiliation{Department of Physics, Dickinson College, Carlisle, PA 17013}

% See the REVTeX documentation for more examples of author and affiliation lists.

\date{\today}

\begin{abstract}
	Our familiar Newton's laws allow determination of both position and velocity of any object precisely. Early nineteenth century saw the birth of quantum mechanics where all measurements must obey Heisenberg's uncertainty principle. Basically, we cannot simultaneously measure with precision, both position and momentum of particles in the microscopic atomic world. A natural extension will be to assume that space becomes fuzzy as we approach the study of early universe. That is, all the components of position cannot be simultaneously measured with precision. Such a space is called non-commutative space. In this article, we study quantum mechanics of hydrogen atom on such a fuzzy space. Particularly, we highlight expected corrections to the hydrogen atom energy spectrum due to non-commutative space. 
\end{abstract}
% AJP requires an abstract for all regular article submissions.
% Abstracts are optional for submissions to the "Notes and Discussions" section.

\maketitle % title page is now complete

\section{Introduction} % Section titles are automatically converted to all-caps.
% Section numbering is automatic.

Let us recall our conventional notion of the trajectories of macroscopic objects $\vec r(t)$ (position $\vec r$ as a function of time) in classical Newtonian mechanics. In the absence of any external forces, $\vec r(t)$ can be obtained from the second order differential equation,
\begin{align}
{d^2 \vec{r}~(t) \over dt^2}=0~,
\end{align}
provided we give both position $\vec r$ and velocity $d\vec r/dt$ at 
initial time $t=t_0$. This concept of simultaneous determination of position and velocity is no longer true once we move to the microscopic atomic world. Particularly, we need a machinery called `quantum mechanics' which is governed by the Heisenberg uncertainty principle.

In the microscopic world, we have the Planck constant $h$ controlling the imprecise measurement of both the observables, position $\vec r$ and momentum $\vec p= m \vec r/dt$. We associate operators to all observables in quantum mechanics. The components of the position and momentum operators obey,
\begin{align}[ \hat{r}_i, \hat{p}_j] =  \hat{r}_i  \hat{p}_j-  \hat{p}_j  \hat{r}_i=  {\iota h \over 2\pi} ~ \delta_{ij}\equiv 
\iota \hbar  \delta_{ij}~,
\end{align}
where the square bracket is called \textit{commutator bracket} as expanded above and $\delta_{ij}$ is the Kronecker delta which is equal to 1 only if $i=j$ and zero otherwise. All the physics of quantum mechanics reduces to classical mechanics when we take the limit
$h \rightarrow 0$.

%Every undergraduate student getting first exposure to the tools of quantum mechanics feels that some concepts are counter-intuitive. For instance, classical notions would not permit particles to have negative kinetic energy but it is allowed quantum mechanically. Interestingly, many counter-intuitive experimental observations like tunneling and discrete energy states of the hydrogen atom require the tools of quantum mechanics.

The next theoretical idea beyond quantum will be to look at physics 
near the big bang singularity which represents the beginning of our expanding universe. Many current research areas like string theory \cite{Seiberg:1999vs} and quantum gravity theories provide us evidence to believe that space near the origin of our universe was fuzzy. As a theoretical idea, which is a natural generalisation of classical to quantum, we introduce a parameter $\theta_{ij}= \frac{1}{2}\epsilon_{ijk} \theta_k$
which is similar to $h$, to govern the fuzziness of space as follows:
\begin{align}
[\hat{x}_i,\hat{x}_j]= \frac{\iota}{2}\sum_{k=1}^{3} \epsilon_{ijk}~\theta_k~,
\end{align}
where $\epsilon_{ijk}$ is the Levi-Civita symbol.
Such a space defined by the above commutation relation is called non-commutative space and $\theta_k$ is the 
\textit{non-commutative (NC) parameter}. In the limit of $\theta_k \rightarrow 0$, we must get back our conventional quantum mechanics on a commutative space. 

It is definitely an interesting exercise to study quantum mechanics of many known systems in the above non-commutative space \cite{Landau, *Smailagic:2001qe}. Here we do so through just a simple modification of taking the coordinate commutator to be non-zero. Our aim is to enable a better understanding of the course content by reviewing the concepts taught in an undergraduate quantum mechanics course in light of this modification. It might also serve to generate exercise problems for students towards getting a hands-on experience in reviewing first order perturbation theory methods, calculating Clebsch-Gordon coefficients, and introduction to spectroscopic notation and Lamb shift.   

In this article, we will initially elaborate the formalism of 
non-commutative quantum mechanics. Then we present a method to solve for energy values in non-commutative space. Finally, we obtain the energy values of the hydrogen atom problem including non-commutative corrections.		
%\subsection*{Subsection heading}
%Subsection text here. Subsection text here. Subsection text here. Subsection text here. Subsection text here. Subsection text here. Subsection text here. 

%%insert keywords separated by comma in the text, exactly aligned to the required margin text.

\section{Non-Commutative Quantum Mechanics (NCQM) - The Formalism}
For a two-dimensional $x-y$ plane, Fig.~1 
gives a pictorial description of fuzziness or non-commutativity 
encoded by parameter $\theta$. 
Such a non-commutative plane is described by,
%Non-commutative Quantum Mechanics shatters this view of space by introducing the commutator between the coordinates as 
\begin{align}
[x   ,  y]  =  \iota \theta~.
\end{align}
%Here, $\theta$ is the \textit{non-commutative (NC) parameter} which is a constant real number determined by experimental data.
As the consequence of a non-zero commutator, there will be an additional uncertainty relation between the two spatial coordinates: 
\begin{align}
\Delta  x ~   \Delta  y \gtrsim  \theta~.
\end{align} 
%At this point we urge the reader to ponder the significance of this result (See figure %something%
%for a hint).\\
\begin{figure}[h!]
	\vskip -12pt
	\centering
	\includegraphics[width=5.5cm, height=5.5cm]{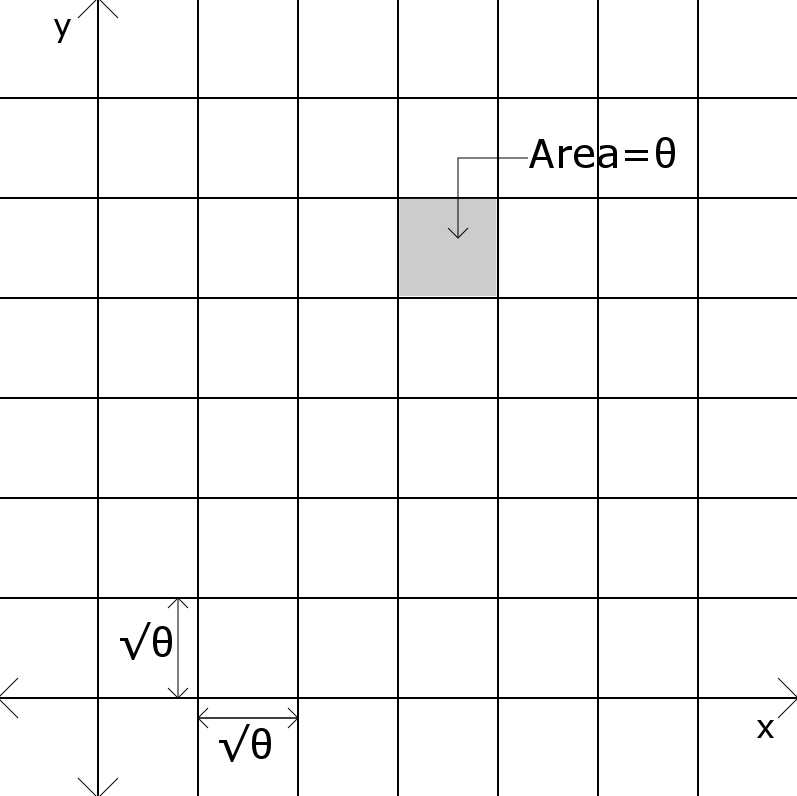}%&
	\caption{The two-dimensional space is divided into cells of area proportional to $\theta$. The uncertainty relation forbids us to resolve space below this area. In n-dimensions, a volume proportional to $(\sqrt{\theta} \hspace{2pt})^n$ will be unresolvable.}\label{NC space}
\end{figure}

The uncertainty as depicted in Fig.~1 means that
%in $x$ and uncertainty in $y$ is less than $\sqrt{\theta}$, then the relation %number%
%is contradicted. This means that 
we cannot be certain about space in an area less than $\theta$.
Hence the position is not given by a point but a fuzzy region of area
$\theta$.
%which further means that the concept of a 'point' as a quantity having zero area has a meaning no longer. \leftHighlight{Non-commutative space is a ``pointless" space!}
In order that all the different particles we observe also be allowed in non-commutative space, the non-commutative parameter is constrained to be $\theta < 10^{-40} \hspace{3pt} m^2$ \cite{doi:10.1142/S0217732319501918}. Our 
present day accelerator experiments cannot probe such length scales
to verify whether space is actually non-commutative or not.
Nevertheless, we would like to revisit hydrogen atom problem 
on non-commutative space and derive the expected non-commutative signatures. We hope future experiments at such length scales will be able to look for the theoretical predictions of non-commutative quantum mechanical systems.
%have experimental evidence at such a length scale exists at present to confirm the non-commutative nature of space. In the later part of this article, we look at a proposed experiment that can calculate the value of the NC parameter, in the process proving the non-commutativity of space. But first, we need to develop a theoretical method to determine the energy values of any given system in NC space.

\section{The Canonical formulation of NCQM}
%\textit{``Physical arguments are but the vision in the painter's dream,
%Mathematical formulation makes the painting complete."}\\\\
A natural generalisation of the non-commutative plane to any $n$-dimensional space will be,
%If there are more than two coordinates, a general NC commutator may be written as :
\begin{align}
[\hat{x}_i,\hat{x}_j] = \iota \theta_{ij} \hspace*{20pt} \text{i,j = 1,2, ... ,n} 
\end{align}
%In the above, $n$ is of course the total dimension of the space, while 
Clearly, the commutator bracket between any two coordinates $x_i$ and $x_j$ is proportional to a constant matrix element $\theta_{ij}$
which must be antisymmetric $\theta_{ij}= - \theta_{ji}$
because $[x_i,x_j]=-[x_j,x_i]$.

% Thus, all $\theta_{ij}$s form a matrix. Let us deduce the form of this matrix, if $i=j$ in the above expression (3),  
%\begin{align}
%[\hat{x}_i,\hat{x}_i] = \hat{x}_i\hat{x}_i - \hat{x}_i\hat{x}_i = 0 = \theta_{ii}
%\end{align}
%Then if we want to find $\theta_{ji}$,
%\begin{align}
%\iota\theta_{ji}= \hat{x}_j\hat{x}_i - \hat{x}_i\hat{x}_j = - %[\hat{x}_i,\hat{x}_j] = - \iota\theta_{ij}
%\end{align}
%Hence we obtain $\theta$ to be a constant antisymmetric matrix.\\
To study quantum mechanics problems in such a non-commutative space,
we will use the well known commutator between
position and momentum but will assume that the momentum components
are simultaneously measurable. That is, the complete set of the commutation relations are,
\begin{align}
\begin{split}
[\hat{x'_i},\hat{x'_j}] &= \iota \theta_{ij}\\
[\hat{p'_i} , \hat{p'_j}] &= 0\\
[\hat{x'_i}, \hat{p'_j}] &= \iota \hbar \delta_{ij}
\end{split}
\end{align}
%The second equation assumes that momentum operators in different coordinates continue to commute (that is, they are unrelated) in NC space. The last expression is a manifestation of the famous uncertainty principle due to Heisenberg\footnote{$\Delta x \Delta y \geq \frac{\hbar}{2}$, the Heisenberg uncertainty relation; is a fundamental relation in quantum mechanics.}. $\delta_{ij}$ is the Kronecker delta\footnote{\textbf{Kronecker delta} : $\delta_{ij}= 1$ if $i=j$ else $\delta_{ij}= 0$ if $i \neq  j $.}. The prime `` \hspace*{3pt} $'$ \hspace*{3pt}" indicates that these operators reside in a different space, the NC space as opposed to un-primed 
%ones below, which reside in the ``ordinary" space.\\
We would like to do a suitable coordinate transformation 
$\hat {x'_i}  \rightarrow \hat {x_i}$ such that
the above commutator relations can be transformed to satisfy,
%The usual relations that we expect a Euclidean space to follow are :
\begin{align}\label{ordinaryrelations}
\begin{split}
[\hat{x}_i,\hat{x}_j] &= 0 \\
[\hat{p}_i , \hat{p}_j] &= 0\\
[\hat{x}_i, \hat{p}_j] &= \iota \hbar \delta_{ij}~.
\end{split} 
\end{align}
The above relations resemble commutative space.
%If we could somehow get the ordinary relations from the NC relations, all the techniques used to tackle problems in ordinary quantum mechanics will be at our disposal. Perhaps the most simple and intuitive way of achieving the above is to make a linear change of variables to move from one space to another. 
The following variable change,
\begin{align}\label{xchange}
\hat{x'_i} & \longrightarrow \hat{x_i} - \sum_{j=1}^{n}\frac{\theta_{ij}}{2\hbar}\hat{p}_j\\
\label{pchange}\hat{p'_i} & \longrightarrow \hat{p}_i 
\end{align}
reproduces the commutative space relations eq.\eqref{ordinaryrelations}. Although this is not the only variable change which gives the correct relations, this form is inspired by another approach to NCQM -namely, the Moyal Product formulation \cite{Gouba:2016iar}.\\

The change in phase space variables eq.\eqref{xchange} and eq.\eqref{pchange} can be used to relate
physics in commutative space to corresponding results in non-commutative space. This makes the evaluation of energy spectrum in problems like hydrogen atom
in non-commutative space straightforward, as will be elaborated in the following section.
%Now the recipe of finding energy is very simple -\\
%`Given a Hamiltonian\mfnote{An operator (which usually is a combination of $\hat{p}$ and $\hat{x}$ operators) whose eigenvalues are precisely the possible energy states of the system.} in NC space, use the above substitution to convert it to ordinary space then get the energy eigenvalues of the system by employing methods already prevalent in ordinary quantum mechanics.' 

\section{Hydrogen Atom in NCQM} 
It is well known that the hydrogen atom, which is a system of an electron and a proton, can be described by an effective one dimensional Hamiltonian, 
\begin{align}
H= {\hat{p}_{r'}^2 \over 
	2 \mu} - { e^2 \over 4\pi\epsilon_0} {1 \over \hat{r}'},
\end{align}
where $\vec r'= \vec r'_e- \vec r'_p$ is the the relative coordinate
with subscripts denoting the electron and proton and 
$\mu=m_e m_p/(m_p+m_e)$ is the reduced mass. Even though
the commutator between the coordinates of electron and proton 
$[\vec r'_e, \vec r'_p]=0$, the commutator between the relative
coordinate components will be,
\begin{align}
[r'_i,r'_j]= 2 \iota \theta_{ij}~\equiv \iota \tilde{\theta}_{ij} \leftrightarrow \iota \theta_{ij}
\end{align}
where we have redefined $\tilde{\theta}$ to $\theta$ for convenience. 
In what follows, we use atomic units such that $4\pi\epsilon_0 = e = m_e = \hbar = 1$, however we track $\hbar$ in the perturbation term to identify its dependence in the result. Suppose we make a coordinate transformation $\vec{r'} \rightarrow \vec r$ and $\vec{p'} \rightarrow \vec p$ as given in eq.\eqref{xchange} and eq.\eqref{pchange},
%\begin{align}\nonumber
%	\hat{x_i} = \hat{x'_i} - \sum_{j=1}^{3}
%	\frac{\theta_{ij}}{2\hbar}\hat{p'}_j \hspace{5pt} ;
%	\hspace{30pt} \hat{p'_i} = \hat{p}_i 
%\end{align}
then the potential $V(|\vec{r'}|)$ will transform as 
%Notice that the only change that occurs is in the potential term of the Hamiltonian,
%%begin{align}
%	V = -\frac{1}{\sqrt{\hat{x'}_i \hat{x'}_i}}
%\end{align}
%where we have omitted the constant terms and the summation sign for brevity. Making the required substitution, we have
\begin{align}
V = - {1 \over \sqrt{\hat{x}'_i \hat{x}'_i}}= - 1/{\sqrt{ (\hat{x_i} - \sum_{j=1}^{3}\frac{{\
				\theta}
			_{ij}}{2\hbar}\hat{p}_j)(\hat{x_i} - \sum_{k=1}^{3}\frac{\theta_{ik}}{2\hbar}\hat{p}_k)                  }}
%		\\
%&= - 1/{\sqrt{ \hat{x_i}\hat{x_i} -  \hat{x_i}\left(\sum_{k=1}^{3}\frac{\theta_{ik}}{2\hbar}\hat{p}_k\right) - \left(\sum_{j=1}^{3}\frac{\theta_{ij}}{2\hbar}\hat{p}_j\right)  \hat{x_i} + \sum_{j,k=1}^{3}\frac{\theta_{ij}\theta_{ik}}{4\hbar^2}\hat{p}_j \hat{p}_k }}\nonumber~,
\end{align}
as the summation variables $j$ and $k$ are equivalent dummy indices the 
above potential will be simplified as, 
\begin{align}
&= - 1/{\sqrt{\hat{x_i}\hat{x_i} -  \sum_{j=1}^{3}\frac{\theta_
			{ij}}{2\hbar}\left(\hat{p}_j  \hat{x_i} + \hat{x_i} \hat{p}_j \right) +  \sum_{j,k=1}^{3}\frac{\theta_{ij}\theta_{ik}}{4\hbar^2}\hat{p}_j \hat{p}_k }}~,
\end{align}
since $[\hat{x}_i, \hat{p}_j] = \iota \hbar \delta_{ij}  \Rightarrow \hat{p}_j \hat{x}_i = \hat{x}_i \hat{p}_j - \iota\hbar\delta_{ij} $, the previous equation simplifies as,
\begin{align}
&= - 1/{\sqrt{ \hat{x_i}\hat{x_i} -  \sum_{j=1}^{3}\frac{\theta_{ij}\left(2 \hat{x_i} \hat{p}_j  - \iota\hbar\delta_{ij}\right)}{2\hbar} +  \sum_{j,k=1}^{3}\frac{\theta_{ij}\theta_{ik}}{4\hbar^2}\hat{p}_j \hat{p}_k }}\\
&=  - 1/{\sqrt{ \hat{x_i}\hat{x_i} -  \sum_{j=1}^{3}\frac{\left(2\theta_{ij} \hat{x_i} \hat{p}_j  - \iota\hbar\theta_{ii}\right)}{2 \hbar} +  \sum_{j,k=1}^{3}\frac{\theta_{ij}\theta_{ik}}{4\hbar^2}\hat{p}_j \hat{p}_k }}~.
\end{align}
We know that $\theta_{ii}$ (antisymmetry property) is zero hence,
\begin{align}
V= - 1/{\sqrt{ \hat{x_i}\hat{x_i} -  \sum_{j=1}^{3}\frac{\theta_{ij} \hat{x_i} \hat{p}_j  } {\hbar} +  \sum_{j,k=1}^{3}\frac{\theta_{ij}\theta_{ik}}{4\hbar^2}\hat{p}_j \hat{p}_k }}~.
\end{align}
Keeping the terms upto $\mathbb{\theta}$ in the potential is sufficient to see the 
signature or correction due to non-commutative space. Hence by
performing the binomial expansion of $V(r)$ we obtain, 
%As it stands, this expression for the potential does not appear to be solvable. Let us make an approximation to neglect terms of higher order in $\theta$. This approximation is valid as $\theta$ is of the order $10^{-40}$ which is very small.
%\begin{align}
%	V&= - \frac{1}{\sqrt{ \hat{x_i}\hat{x_i} -  \sum_{j=1}^{n}\frac{\theta_{ij}}{\hbar} \hat{x_i} \hat{p}_j }}\\
%%Doubt  	&= - \left(\hat{x_i}\hat{x_i} -  \sum_{j=1}^{n}\frac{\theta_{ij}}{\hbar} %\hat{x_i} \hat{p}_j \right)^{-\frac{1}{2}}\\
%&= - \left(\hat{x_i}\hat{x_i}\right)^{-\frac{1}{2}} \left( 1 - \frac{1}%{\hat{x_i}\hat{x_i}} \sum_{j=1}^{n}\frac{\theta_{ij} \hat{x_i} \hat{p}_j}{ \hbar}\right)^{-\frac{1}{2}}
%\end{align}
%Now we take the binomial expansion of the term in bracket while keeping in %mind that it is valid only when $|\sum_{j=1}^{n}\frac{\theta_{ij}}{\hbar} \frac{\hat{x_i} \hat{p}_j}{\hat{x_i}\hat{x_i}}|  << 1 $. Since $\theta$ is small, this is true unless $\hat{x_i}\hat{x_i}$ is very small. In other words, the radial distance $r= \sqrt{\hat{x_i}\hat{x_i}}$ must not approach zero.
\begin{align}\label{r=0}
V &= - \left(\hat{x_t}\hat{x_t}\right)^{-\frac{1}{2}} \left( 1 + \frac{1}{2} \frac{1}{\hat{x_t}\hat{x_t}}\sum_{i,j=1}^{3}\frac{\theta_{ij}\hat{x_i} \hat{p}_j}{\hbar}  + \text{Order ($\theta^2$)} \right)\\ 
&= -\frac{1}{r} - \frac{1}{2\hbar r^3} \sum_{i,j=1}^{3}\theta_{ij}\hat{x_i} \hat{p}_j~,
\end{align}
where $r=\sqrt{\hat{x_t}\hat {x_t}}$ must not be zero. Recall, 
$\theta_{ij} = \frac{1}{2}\sum_{k}^{}\epsilon_{ijk}\theta_{k}$. Hence the potential can be 
simplified as, 
\begin{align}
V = -\frac{1}{r} - \frac{1}{4\hbar r^3} \sum_{i,j,k=1}^{3}\epsilon_{ijk}\theta_{k} \hat{x_i} \hat{p}_j~.
\end{align} 
This expression shows the additional term is dependent on angular momentum 
operator 
%From here, we recognise the familiar form of the angular momentum operator in the $k$ direction, 
$\hat{L}_k =\sum_{i,j=1}^{3} \epsilon_{ijk} \hat{x_i} \hat{p}_j $, 
\begin{align}
V&= -\frac{1}{r} - \frac{1}{4\hbar r^3} \sum_{k=1}^{3}\hat{L}_k \theta_k\\
V&=  -\frac{1}{r} - \frac{1}{4\hbar r^3} \hat{\vec{L}} \cdot \vec{\theta}~.
\end{align}
Thus we have obtained a perturbation term to the ordinary $1/r$ potential for the hydrogen atom. Taking $\vec{\theta} = | \vec{\theta}| \hat{n}$ and choosing the direction $\hat{n}$ along z-axis implies the perturbation in this choice of frame is,   $- \frac{1}{4\hbar r^3} \theta_z \hat{L}_z$. The energy spectrum can now be found using first order perturbation,
\begin{align}
\Delta E = \bra{n,l,m} - \frac{ \theta_z \hat{L}_z }{4\hbar r^3}\ket{n,l,m}~.
\end{align}
As $\hat{L}_z$ commutes with the perturbation free Hamiltonian, the hydrogen atom eigenstates $\ket{n,l,m}$ are eigenstates of $\hat{L}_z$ too, satisfying
\begin{align}
\hat{L}_z \ket{n,l,m} = m \hbar \ket{n,l,m}~.
\end{align}  
Thus the perturbation expression simplifies to,
\begin{align}
\Delta E = - \frac{ \theta_z  m }{4} \bra{n,l,m} \frac{1}{r^3}\ket{n,l,m}~.
\end{align} 
The expectation value for $1/r^3$ can be obtained by a beautiful trick \cite{Shankar},
\begin{align}
\bra{n,l,m} \frac{1}{r^3} \ket{n,l,m} = \frac{1}{n^3l(l+1/2)(l+1)} ~.
\end{align}
For obtaining the expectation value, we need to integrate $1/r^3$ from zero to infinity, but our earlier approximation eq.\eqref{r=0} that the radial distance must not approach zero is violated. This does not pose a problem as for $l \neq 0$, the hydrogen wave-function tends to zero as $r$ tends to zero, making the integral zero there. The $l=0$ states on the other hand have $m=0$ and hence do not contribute in the perturbation expression, which now is
\begin{align}
\Delta E = -\frac{\theta_z m}{4n^3l(l+1/2)(l+1)}~.
\end{align}
Essentially the problem that we began with, namely, to obtain the energy eigenvalues of the hydrogen atom has been solved, at least to first order in perturbation theory. 

In all of the above, we have used $n,l$ and $m$ quantum numbers to denote the states of hydrogen atom. But it is known through experiments, that for hydrogen atom the \textit{total angular momentum} is what is conserved. Accordingly, a more complete treatment would demand that we consider both the orbital as well as spin quantum numbers.
This can be done easily by introducing in the place of $\ket{n,l,m}$ the states $\ket{n,j,j_z}$ where $j = l+s$ is the total angular momentum quantum number.\\
Returning back to the first order perturbation expression and performing the appropriate replacement of eigenstates,
\begin{align}
\Delta E = \bra{n,j,j_z} - \frac{ \theta_z \hat{L}_z }{4\hbar r^3}\ket{n,j,j_z} ,
\end{align}
Now consider only,
\begin{align}
\bra{n,j,j_z} \hat{L}_z \ket{n,j,j_z}
\end{align}
By completeness condition, $\hat{\mathbb{I}}=\ket{n;l,l_z;s,s_z}\bra{n;l,l_z;s,s_z}$, dropping the $n$ for brevity we can write,
\begin{align}
&=  \bra{j,j_z} \hat{L}_z \left(\sum_{l_z,s_z}\ket{l,l_z;s,s_z}\bra{l,l_z;s,s_z}\right) \ket{j,j_z}\\
&= \sum_{l_z,s_z} \bra{j,j_z} \hat{L}_z \ket{l,l_z;s,s_z}\braket{l,l_z;s,s_z|j,j_z}\\
&= \sum_{l_z,s_z} l_z\hbar \braket{j,j_z|l,l_z;s,s_z}\braket{l,l_z;s,s_z|j,j_z}
\end{align}
Noting that $\braket{j,j_z|l,l_z;s,s_z}$ and $\braket{l,l_z;s,s_z|j,j_z}$ are Clebsch-Gordan (CG) coefficients which are equal,
\begin{align}\label{l=0}
&= \sum_{l_z,s_z} l_z \hbar \left| \braket{l,l_z;s,s_z|j,j_z} \right|^2~.
\end{align}
Here we see that if $l=0$, $l_z$ has only one value, $l_z = 0$, hence the expression is zero for $l=0$ states. In the appendix we show how to find a general Clebsch-Gordan expression when $s=1/2$ \cite{Shankar2}. Substituting the values of CG coefficients and performing the summation, the reader can verify that,
\[
\bra{n,j,j_z}\hat{L}_z \ket{n,j,j_z} =
\begin{cases}
\left(1 -  \frac{1}{2l+1}\right) j_z \hbar \hspace{3pt} &\text{for} \hspace{3pt} j = l + 1/2\\
\left(1 +  \frac{1}{2l+1}\right) j_z \hbar \hspace{3pt} &\text{for} \hspace{3pt} j = l - 1/2\\ 
\end{cases}
\]
With this in mind, the first order energy difference $\Delta E$ simplifies to,
\begin{align}
\Delta E = -\frac{\theta_z}{4} \left(1\mp \frac{1}{2l+1}\right) j_z 
\bra{n,l',j,j'_z} \frac{1}{r^3} \ket{n,l,j,j_z}~.
\end{align}
Once again, for obtaining the expectation value, we seemingly violate our earlier approximation (15) that the radial distance must not approach zero. But here too it goes through as for $l \neq 0$, the hydrogen wave-function tends to zero as $r$ tends to zero, making the integral zero there. The $l=0$ states on the other hand are seen eq.\eqref{l=0} to undergo no correction at first order. 
We have thus obtained the energy spectrum for hydrogen atom in non-commutative space, this time incorporating spin, as:
\begin{align}
\Delta E = - \frac{\theta_z j_z}{4} \left(1\mp \frac{1}{2l+1}\right) \left(\frac{1}{n^3l(l+1/2)(l+1)}\right) ~.
\end{align}
This result was first derived using quantum field theory arguments in ref.\cite{PhysRevLett.86.2716}.

%Before interpreting what we have found let us first briefly take a detour to the year 1947, when Willis Lamb and Robert Retherford %someone 
%performed their famous experiment for obtaining the hydrogen energy levels.

\section{Lamb's Shift and Non-commutative correction}
%When applied to the system of hydrogen atom, due to spin effects some \textit{states}\footnote{A \textit{state} is similar to a trajectory of a particle in Newtonian mechanics, more precisely, a state gives the probability distribution in space that the particle should lie in.}have same energy while others differed in their energies. This is known as \textbf{splitting of energies}, when two \textit{states} which were earlier believed to have the same energy now are split into different energy values.
%
%A serious blow came when in 1947, Willis Lamb and Robert Retherford experimentally found a splitting of energies between two of the hydrogen atom states which were predicted to have the same energy by the Dirac equation. 
%Before the year 1947, the prevalent equation of state evolution which could combine relativistic effects with the quantum nature of particles was the Dirac equation. This equation successfully describes the spin quantum number and its consequences. 
For the more methodically minded reader, the ad-hoc inclusion of spin quantum number in the treatment above might seem a little incoherent. This concern can safely be abandoned as spin is known to come out naturally and rather miraculously by incorporating relativity into quantum mechanics via the Dirac equation.\\ 
%Before the year 1947, Dirac equation was the cornerstone of modern physics. It was ``the" equation to start with in order to get energy and evolution of any system. 
%It is similar to  Newton's equations which determine the evolution of a particle's trajectory, the difference being the Dirac equation is also valid when the particle is microscopic and moves with velocity comparable to the speed of light.
In the case of hydrogen atom for example the energy spectrum through Dirac equation is found to depend on $n$ and $j$,
\begin{align}\nonumber
j= &\{l + s, l+s-1,...\\
&...\text{decreasing in steps of one until}, \left| l-s \right| \} 
\end{align}  
where the $j$ indicates coupled angular momentum and the energy is different for each $j$. A striking consequence of this result is that the energies of $\ket{n=2, l=1, s=1/2}$, which are split according to,
\[ j = \begin{cases} 
1 + (1/2)  = 3/2 & \rightarrow {^2P}_{3/2}\\
1 + (1/2) -1 = 1/2 & \rightarrow {^2P}_{1/2} \\
\end{cases}
\] 
and $\ket{n=2, l=0, s= 1/2}$ having, 
\begin{align}\nonumber
j = 0 + (1/2)  = 1/2 ~ \rightarrow {^2S}_{1/2} 
\end{align}
are equal when $j$ for both is $1/2$ while $n=2$ is same for both states. In the above, the notation following the arrows represents coupled states and is known as the `spectroscopic' or `term symbol' notation. This notation is expressed as $^{2s+1}l_j$.\\
The prediction of equal energies of ${^2P}_{1/2}$ and ${^2S}_{1/2}$ states was shattered in 1947, through experiments conducted by Lamb and Retherford who found that there was a difference in energies between these states. The difference became famous as `Lamb Shift' for which Willis E. Lamb was awarded the 1955 Nobel prize in physics, ``\textit{for his discoveries concerning the fine structure of the hydrogen spectrum}".
%Thereafter physicists all around the world plunged into reformulating quantum mechanics to explain this failure of the Dirac equation. Then, after many attempts, arose `Quantum Electrodynamics', a quantum mechanical formulation that takes into account the quantum nature of electric and magnetic fields along with the quantum nature of particles. Indeed, the `Lamb Shift' is one of the pillars on which renormalisation theory\footnote{In a crude sense, renormalisation provides particles with a ``bare" charge or mass which is then modified by the presence of fields into the experimentally observed values.} stands.\\
%The difference between the previous two energy states can be crucially attributed to the fact that the energy levels depended on $n$, $j$ (and hence $l$) quantum numbers. Let us now come back to our non-commutative space and investigate the first order correction term to the energy we obtained in the previous section (49) ,

Let us investigate what happens to the above two states in the NC space energy spectrum we derived by means of an ad-hoc introduction of spin,
\begin{align}\label{NCspectrum}
\Delta E = - \frac{\theta_z j_z}{4} \left(1\mp \frac{1}{2l+1}\right) \left(\frac{1}{n^3l(l+1/2)(l+1)}\right) ~.
\end{align}             
We can readily see that the energy is in fact dependent on both $l$ and $n$ quantum numbers thus giving a shift between the aforementioned states. But there is more! A $j_z$ term is also present which tells us that the energy is further split according to, 
\begin{align}
j_z = \{ -j , -j+1, ..., \text{increasing in steps of one till}, +j \}   
\end{align}
Specifically in our case, the $l=0$ states are shown to have no correction to first order eq.\eqref{l=0}, so $^2S_{1/2}$ level is not changed. The $^2P_{1/2}$ level on the other hand undergoes a correction as,
\[ ^2P_{j = 1/2} = \begin{cases} 
^2P_{-1/2}  & \text{for} \hspace{10pt}  j_z = -j = -1/2 \\
^2P_{+1/2}& \text{for} \hspace{10pt}   j_z = -j+1 = +1/2 = +j \\
\end{cases}
\] 
Thus, the `Lamb Shift' itself is split into two lines corresponding to different $j_z$. (See Fig.2. As a cautionary note, although we have placed NCQM after QED in the figure, while deriving the shift we have not used any field theoretical arguments.)
\begin{figure}[h!]
	\caption{Hydrogen atom energy levels according to different theories. `QED' stands for Quantum Electrodynamics while `NCQM' is Non-commutative Quantum Mechanics. The $^2P_{1/2}$ level splits into $^2P_{-1/2}$ and $^2P_{+1/2}$ when we consider the non-commutativity of space.} \label{H atom spectrum}
%	\vskip -12pt	
	\centering
	\includegraphics[width= 8.5cm, height=4.5cm]{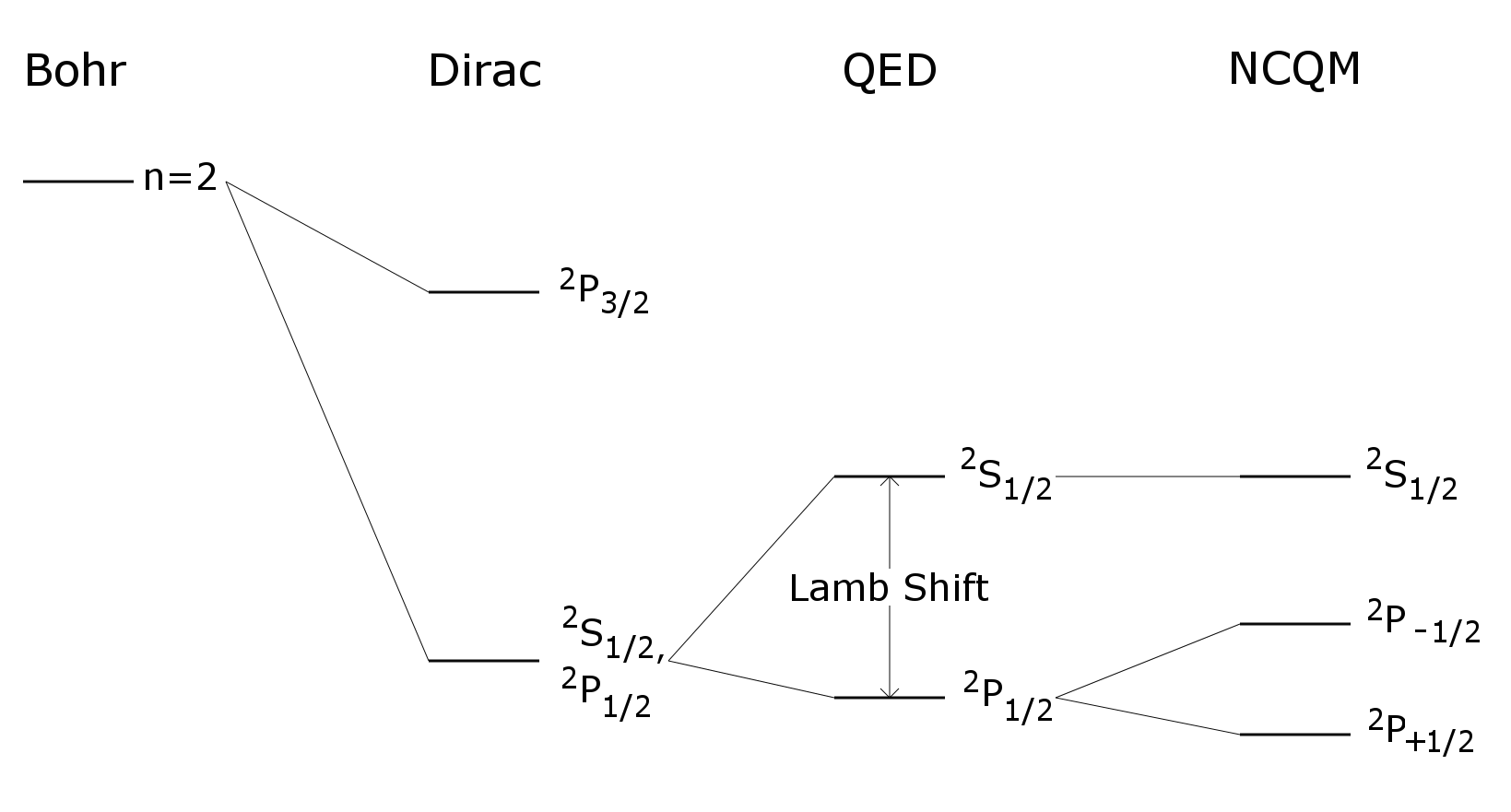}		
	%&
\end{figure}

Although we started out with undergraduate level quantum mechanics we have now been led to the level of Quantum Electrodynamics, a theory which lies at the frontier of physics today! The most monumental success of quantum electrodynamics is its explanation for the Lamb Shift in hydrogen atom. It is indeed one of the pillars on which the theory stands. Our investigations suggest that we can expect even finer structure to the Lamb Shift!\\
The next question that we have to ask is that why is this splitting of Lamb Shift not detected, as surely many modern experiments would have verified the Lamb Shift with increasing precision. The answer to this also lies in our energy expression, the presence of $\theta_z$ term tells us that this splitting is of the order of $\theta$ which is very small for the present technology to detect. Indeed, as we had pointed out earlier, this splitting can be used as a confirmation that the space we live in is actually non-commutative. Not only this, but also a measurement of the amount of splitting can estimate the value of the non-commutativity parameter directly by using eq.\eqref{NCspectrum}.

\section{Conclusion}
In this article, we have taken a tour through the world of non-commutating space coordinates. In doing so, we have studied a way to incorporate non-commutativity into our usual framework of quantum mechanics and then applied this method to get a splitting in the Lamb shift of the hydrogen atom. Although it is as yet unknown whether our space is non-commutative, we hope to have impressed upon the reader that even a simple and generally assumed notion like the commutator for space coordinates being zero, when inspected more thoroughly, can reveal some new and exciting physics.

\section*{Acknowledgements}
We would like to thank Toby Wiseman of Theoretical Physics Group at Imperial College London for helpful comments on the manuscript. PSG acknowledges support from the Rajarshi Shahu Maharaj foreign education scholarship 2018-19 of the Government of Maharashtra State, India.

\appendix*   % Omit the * if there's more than one appendix.
\section*{Appendix}
The total angular momentum operator is given by,
\begin{align}
\textbf{J} = \textbf{L} + \textbf{S} 
\end{align}
Squaring, we find
\begin{align}
2\textbf{L} \cdot \textbf{S}  = \textbf{J}^2  - \textbf{L}^2  - \textbf{S}^2 = 2L_zS_z + L_{+}S_{-} + L_{-}S_{+}
\end{align}
where $A_{\pm} = A_x \pm \iota A_y$. The eigenstates of $\textbf{J}^2$ are same as those of $\textbf{L}\cdot \textbf{S}$ as can be proved by obtaining commutator between them to be vanishing. Therefore the eigenvalue equation for $2\textbf{L}\cdot \textbf{S}$ can be written as,
\begin{align}
(2L_zS_z + L_{+}S_{-} + L_{-}S_{+}) \ket{j,j_z} = \lambda \ket{j,j_z}
\end{align}
Pre-multiplying both sides by $\bra{l,l_z;s,s_z}$,
\begin{align}\nonumber
&\bra{l,l_z;s,s_z} (2L_zS_z + L_{+}S_{-} + L_{-}S_{+}) \ket{j,j_z}\\
 &=  \lambda \braket{l,l_z;s,s_z|j,j_z}
\end{align}
The term on the right hand side is recognised as the CG coefficient. Now there are two equations for the two possible values of $s_z$. For $s_z = -1/2 $, $l_z= j_z +1/2$ and for $s_z = +1/2$, $l_z = j_z -1/2$. In terms of the CG coefficients $a = \braket{l, j_z +1/2;1/2, -1/2|j,j_z}$ and $b = \braket{l,j_z - 1/2;1/2, +1/2|j,j_z}$, and using the operator rules for $A_z, A_{+}$ and $A_{-}$ operators, the above two equations can be written as,  
%\begin{align}
% \bra{l,l_z;\frac{1}{2}, -\frac{1}{2}}(2L_zS_z + L_{+}S_{-} + L_{-}S_{+}) \ket{j,l_z - \frac{1}{2}} = \lambda \braket{l,l_z;\frac{1}{2}, -\frac{1}{2}|j,l_z - \frac{1}{2}}
%\end{align}
%and for $s_z =+1/2$, $j_z= l_z + 1/2$,
%\begin{align}
%\bra{l,l_z;1/2, +1/2}(2L_zS_z + L_{+}S_{-} + L_{-}S_{+}) \ket{j,l_z +1/2} = \lambda \braket{l,l_z;1/2, +1/2|j,l_z+1/2}
%\end{align}
\begin{align}
\sqrt{(l - j_z + 1/2) (l+j_z+1/2)}\hspace{3pt}b - (j_z + 1/2) \hspace{3pt}a &= \lambda \hspace{3pt} a\\
(j_z - 1/2) \hspace{3pt}b +  \sqrt{(l - j_z + 1/2) (l+j_z+1/2)} \hspace{3pt}a &= \lambda\hspace{3pt} b
\end{align} 
Solving the above two equations in terms of $l$ and $j_z$ we have,
\[
b/a = \begin{cases}
\sqrt{\frac{l+j_z +1/2}{l-j_z +1/2}} \hspace{10pt} &\text{for} \hspace{4pt} \lambda = l\\
-\sqrt{\frac{l-j_z +1/2}{l+j_z +1/2}} \hspace{10pt} &\text{for} \hspace{4pt} \lambda = -l-1
\end{cases}
\]
The coefficients are required to satisfy $a^2 + b^2 = 1$. Also the standard sign convention for CG coefficients dictates that we take sign for $a$ to be positive. Keeping in mind the orthogonality between CG coefficients, the values for $a$ and $b$ can be unambiguously obtained as shown below,
\begin{table}[h]
	\centering
	\setlength{\tabcolsep}{0.5em} % for the horizontal padding
	{\renewcommand{\arraystretch}{1.75}% for the vertical padding
		\begin{tabular}{lll}
			\hline
			\multicolumn{3}{c}{CG Coefficients for s=1/2}           \\ \hline
			\multicolumn{1}{l|}{}          & $s_z= 1/2$ & $s_z =-1/2$ \\ \hline
			\multicolumn{1}{l|}{j = l+1/2} & $\sqrt{\frac{l + j_z + 1/2}{2 l + 1}} $          &  $ \sqrt{\frac{l - j_z + 1/2}{2 l + 1}}$         \\
			\multicolumn{1}{l|}{j= l-1/2}  &  -$\sqrt{\frac{l - j_z + 1/2}{2 l + 1}} $          &  $\sqrt{\frac{l + j_z + 1/2}{2 l + 1}} $ \\ \hline         
		\end{tabular}
	}
\end{table} 

\bibliography{NCHydrobib}

\end{document}